\documentclass[a4paper,12pt]{article}
\pdfoutput=1
\usepackage{amssymb,amsmath,bm}
\usepackage{color}
\usepackage{slashed}
\usepackage{cite}
\usepackage[latin1]{inputenc}
\usepackage{amsfonts}
\usepackage{lscape}
\usepackage{amsthm}
\usepackage{booktabs}
\usepackage{array}
\usepackage{rotating}
\usepackage{float}
\usepackage{multirow}
\usepackage{graphicx,rotating,amsmath,multirow,xcolor,colortbl,xcolor}
\usepackage{hyperref,pdflscape,slashed}
\definecolor{rosso}{cmyk}{0,1,1,0.4}
\definecolor{rossos}{cmyk}{0,1,1,0.55}
\definecolor{rossoc}{cmyk}{0,1,1,0.2}
\definecolor{blu}{cmyk}{1,1,0,0.3}
\definecolor{blus}{cmyk}{1,1,0,0.6}
\definecolor{bluc}{cmyk}{1,1,0,0.1}
\definecolor{verde}{cmyk}{0.92,0,0.59,0.25}
\definecolor{verdec}{cmyk}{0.92,0,0.59,0.15}
\definecolor{verdes}{cmyk}{0.92,0,0.59,0.4}
\definecolor{Gray}{gray}{0.95}

\font\tenrsfs=rsfs10 at 12pt
\font\sevenrsfs=rsfs7
\font\fiversfs=rsfs5
\newfam\rsfsfam
\textfont\rsfsfam=\tenrsfs
\scriptfont\rsfsfam=\sevenrsfs
\scriptscriptfont\rsfsfam=\fiversfs
\def\mathscr#1{{\fam\rsfsfam\relax#1}}

\usepackage[small,bf]{caption}
\hypersetup{colorlinks,bookmarksopen,bookmarksnumbered,
linkcolor=blus,pdfstartview=FitH,urlcolor=rossos,citecolor=verde}

\newcommand{\be}{\begin{equation}}
\newcommand{\ee}{\end{equation}}
\newcommand{\bea}{\begin{eqnarray}}
\newcommand{\eea}{\end{eqnarray}}
\newcommand{\beq}{\begin{equation}}
\newcommand{\eeq}{\end{equation}}
\newcommand{\beqa}{\begin{eqnarray}}
\newcommand{\eeqa}{\end{eqnarray}}

\newcommand{\R}{\mathbb{R}}
\newcommand{\Z}{\mathbb{Z}}

\def\eq#1{eq.~(\ref{#1})}

\def\sect#1{sect.~\ref{#1}}

\def\Eq#1{eq.~(\ref{#1})}

\def\Sec#1{sect.~\ref{#1}}
\begin{document}

{\hfill CERN-TH-2017-108}

\vspace{1cm}

\begin{center}
\boldmath

{\textbf{\LARGE Comment on ``Disassembling\\[2mm]
the Clockwork Mechanism"
}}
\unboldmath

\bigskip

\vspace{0.4 truecm}

{\bf Gian F. Giudice} and {\bf Matthew McCullough}
 \\[5mm]

{\it CERN, Theoretical Physics Department, Geneva, Switzerland}\\[2mm]

\vspace{0.8cm}

{\bf Abstract }

\vspace{0.2cm}
%\begin{quote}
{\it We respond to the criticism raised in the paper arXiv:1704.07831.}
%\end{quote}
\end{center}
\thispagestyle{empty}
%\vfill

%\newpage
\vspace{1.5cm}
%\tableofcontents
%\newpage

We wish to respond to the criticism of \cite{Giudice:2016yja} made by the authors of \cite{Craig:2017cda}. We start by reviewing in sect.~1 the use of the term `clockwork' employed in \cite{Giudice:2016yja} and 
then, in the following sections,
we address each objection raised in \cite{Craig:2017cda}. We summarise our point of view in \sect{sec:summary}.

\section{Revisiting the clockwork}
\label{sec:iflifegivesyoulemmas}
The clockwork is a theory in which the phases of $N\! +\! 1$ different $\text{U}(1)_j$  symmetries ($j=0,...,N$) are `clockworked', meaning that the original symmetry is explicitly broken in such a way to preserve only a single $\text{U}(1)$, under which the original $\text{U}(1)_j$ factors rotate synchronously with phases $\Delta \theta_{j} = q \Delta \theta_{j+1}$, where $q$ is the clockworking factor. For a chain of spontaneously broken global symmetries this leads to a single Goldstone boson zero mode, whose properties are completely fixed by this symmetry pattern.  As a result, the zero mode is exponentially localised at one end of the chain, by a factor $q^N$ relative to the other end. These theories were introduced in \cite{Choi:2015fiu,Kaplan:2015fuy} and generalised in various ways in \cite{Giudice:2016yja}. To avoid committing to any particular UV-completion, and to allow for general non-integer $q$, the approach followed in \cite{Giudice:2016yja} was based on effective field theories (EFT), and thus focussed on their defining elements: symmetry and field content.

The general EFT of a clockwork for a spontaneously broken $\text{U}(1)^{N+1}$ symmetry contains
$N\! +\! 1$ Goldstones with a Lagrangian
\beq
{\mathcal L} = -\frac{f^2}{2}\, \sum_{j=0}^N \partial_\mu U_j^\dagger\, \partial^\mu U_j + {\rm higher~derivative~terms}.
\label{lag1}
\eeq
Here $U_j = e^{i \pi_j/f}$, where $\pi_j$ is the Goldstone of the $\text{U}(1)_j$ factor, and $f$ denotes the scale of spontaneous symmetry breaking. Fields have been normalised to have canonical kinetic terms. The clockworking is introduced by a potential that explicitly breaks the global symmetry, up to a synchronous field phase rotation with $\Delta \theta_{j} = q \Delta \theta_{j+1}$. In the effective theory, this pattern is achieved with a potential given by a generic function of the combinations $U_j^\dagger\, U_{j+1}^q$. Imposing that interactions occur only between neighbouring sites, we can write the potential as a general power expansion of the form
\beq
V=-\frac{m^2\, f^2}{2} \sum_{j=0}^{N-1} \sum_{z=0}^\infty  c_{jz} (U_j^\dagger\, U_{j+1}^q)^z +{\rm h.c.} =
- m^2\, f^2 \sum_{j=0}^{N-1} \sum_{z=0}^\infty  c_{jz} \cos\left[ \frac{z}{f}(\pi_j -q\pi_{j+1})\right] ~.
\label{lag2}
\eeq
The coefficients $c_{jz}$ encode all model dependence or, in other words, the structure of the UV theory.  Throughout this work we will be interested in the localisation of the fields, which only depends on the quadratic part of the action. Thus it is useful to expand this action in fields, such that \Eq{lag2} may equivalently be written as
\beq
{\mathcal L} = -\frac{1}{2}\, \sum_{j=0}^N \partial_\mu \pi_j \, \partial^\mu \pi_j- \frac{m^2}{2}  \sum_{j=0}^{N-1}  (\pi_j - q \pi_{j+1})^2 + {\mathcal L}_{\rm int} (\pi_j -q\pi_{j+1}) ~,
\label{eq:quad}
\eeq
where we have imposed site-independent neighbouring mass terms, and the model-dependent interaction terms are
\be
{\mathcal L}_{\rm int} (\pi_j -q\pi_{j+1}) = m^2\, f^2  \sum_{j=0}^{N-1} \sum_{z=2}^\infty c'_{jz}  \left (\frac{\pi_j - q \pi_{j+1}}{f}\right)^{2 z}  ~.
\ee
This is a generalisation of the models in \cite{Choi:2015fiu,Kaplan:2015fuy}, which are included as special cases.

Note that in any theory of this class there is a single continuous symmetry, $U(1)_{CW}$, under which the fields $U_j$ carry charges $q^{-j}$.  The coefficients $c_{jz}$, treated as spurions, are invariant under this symmetry, and hence the properties of the associated Goldstone boson do not depend on them.  Thus, purely from the clockwork symmetry $U(1)_{CW}$, we can derive the properties of the Goldstone boson.  

The $U(1)_{CW}$ symmetry acts nonlinearly on the fields, as a simultaneous shift $\pi_j \to \pi_j +q^{-j}\alpha$, where $\alpha$ is a transformation parameter. This symmetry dictates that the zero mode $a_0$ is such that
\beq
m^2_{a_0} =0\, ,~~~ a_0 =\sum_{j=0}^N O_{j0}\, \pi_j \,~~~
O_{j 0} = \frac{{\cal N}_0}{q^j}  ~~,~~ {\cal N}_0 \equiv \sqrt{\frac{q^2-1}{q^2-q^{-2N}}}  ~.
\label{mass}
\eeq
The form of $O_{j 0}$ exhibits the exponential localisation of the zero mode, with the corresponding suppression of its component at the site $N$ by a factor $q^{-N}$. The quadratic terms in \eq{eq:quad} determine the characteristic mass spectrum of the clockwork gears, $m_{a_k}^2 =m^2 (q^2+1-2q\cos \frac{k\pi}{N\!+\!1})$, and their compositions.

A crucial consequence of the zero mode localisation lies in its coupling to external states.  For example, if a single $\pi_\ell$ couples to a $\text{U}(1)_\ell$ current, such as fermion or Chern-Simons currents, with charge $Q_\ell$, then the massless mode acquires a coupling with a suppression that depends exponentially on the position
\be
Q_\ell \partial_\mu \pi_\ell J^\mu_\ell \to {\cal N}_0 \frac{Q_\ell}{q^\ell} \partial_\mu a_0 J^\mu_\ell ~.
\ee
Importantly, the coupling of the $k$-th massive mode is not suppressed by $q^\ell$, meaning that the effective coupling of the zero mode is suppressed relative to the coupling of the gear by a factor
\beq
\frac{O_{\ell0}}{O_{\ell k}} \sim  \frac{{\cal N}_0}{q^\ell\, {\cal N}_k} ~,
\label{rapporto}
\eeq
where ${\cal N}_k$ is $\mathcal{O}(1)$, since the gears are uniformly spread along different sites.
This ratio is a physical observable, which is well-defined in the EFT and measures the localisation of the zero mode. 
As the true cutoff of the interaction is $f/Q_\ell$, this ratio expresses the hierarchy between the interaction scale of the zero mode and the cutoff scale.

This hierarchy of scales, generated by the clockwork symmetry, is independent of any model-dependent features, such as the size of $Q_j$ or the $c_{j z}$'s.   This feature formed the basis of applications of the clockwork proposed in \cite{Giudice:2016yja} and makes clockworking appealing for a wide range of model-building and phenomenological settings.  All models in \cite{Giudice:2016yja} exhibit this clockwork symmetry and the corresponding features.

\section{On clockwork applications}
\label{secdef}
The goal of \cite{Giudice:2016yja} was to find applications of the clockworked symmetry in the context of EFT model building. These models were expressly designed to tackle various phenomenological questions relevant to the low-energy domain, but were not intended to address issues such as transplanckian field excursions or the Weak Gravity Conjecture (WGC)~\cite{ArkaniHamed:2006dz}. They focused on setups in which the clockwork system is coupled to an external sector at a {\it single} site: the last site of the chain (in the discrete version) or the brane at the boundary of the 5D space (in the continuum version). The generalisations of the models in \cite{Choi:2015fiu,Kaplan:2015fuy} were constructed in such a way to retain the original clockwork symmetry structure and, as a consequence, its ensuing properties. As recalled in \sect{sec:iflifegivesyoulemmas}, these properties are:  the localisation of the zero mode, the characteristic gear mass spectrum, the aforementioned hierarchy between the zero-mode interaction scale and the cutoff. 

The definition of clockwork used in \cite{Craig:2017cda} is centred on the existence of zero-mode couplings exponentially varying at different sites, in the absence of exponential hierarchies in fundamental parameters. This definition focuses on the properties of the zero-mode alone (no reference to the gears), as the clockwork system is coupled to external sectors on different sites. The definition in \cite{Craig:2017cda} is useful to address various UV aspects related to transplanckian field excursions or to the WGC, which were indeed some of the original motivations of the work in \cite{Choi:2015fiu,Kaplan:2015fuy}. However, the definition is not befitting to capture the clockwork properties in the EFT domain that was considered in \cite{Giudice:2016yja}, as it is based on an assumption about the couplings between the clockwork and the external sector at {\it all} sites. The coupling profile along sites is a UV feature, which is model-dependent and not dictated by the clockwork symmetry alone. On the contrary, the ratio between the couplings of the zero mode and the gears (or, put another way, the ratio between the cutoff scale and the zero-mode interaction scale) is calculable in the EFT and is a direct consequence of clockworking, independently of any specific model implementation. This ratio changes exponentially along sites: it is of order unity at one end of the chain and exponentially small at the opposite end. With the use of their definition, the authors of \cite{Craig:2017cda} exclude models that share the symmetry structure characteristic of the clockwork and have appealing properties for model-building and phenomenological applications. 

The difference between the two different approaches lies in an assumption on the charges of the clockwork system to external sectors. In  \cite{Craig:2017cda} it is argued that these charges are such that $Q_j=1$ ($Q^{CW}_j = q^{-j}$). This assumption is essentially related to locality, in the sense of the dimensionality of the operator that couples the clockwork to an external state, and is valid for certain UV completions, for example involving the complex scalar with the radial mode included ($U_j \to \Phi_j$), as in the models of \cite{Choi:2015fiu,Kaplan:2015fuy}. Thus, according to the definition adopted in \cite{Craig:2017cda}, if a UV model generates the clockwork EFT, eqs.~(\ref{lag1})--(\ref{lag2}), but also a non-uniform pattern of $Q_j$, then it is deemed `not clockwork'.  As the zero mode is protected by the non-linearly realised $U(1)_{CW}$ symmetry, a hierarchy of couplings to external states necessarily requires a hierarchy of charges under $U(1)_{CW}$, and hence charge quantisation.  This automatically restricts to Abelian theories, not by proof but by definition. 

The ingredients of the EFT in eqs.~(\ref{lag1})--(\ref{lag2}) are the clockwork symmetry, nearest-neighbour interactions, and site-independent mass terms.   The power-counting for the approximate symmetries may be motivated by spurion analysis.  In the EFT the coupling of the field $\pi_j$ to external fields must respect the shift symmetry, but that is all.  Different UV completions may motivate different magnitudes of couplings. In the clockwork EFT, when going to the mass eigenstate basis, the zero mode inherits from $\pi_j$ a coupling that is exponentially suppressed relative to the gears, and to the cutoff of that interaction.  This is the only model-independent prediction of an effective clockwork theory and this is what guided the work in \cite{Giudice:2016yja}.

Not only does the analysis in \cite{Giudice:2016yja} make no assumption on the external charge profile, but it restricts considerations to models where the coupling is localised on a single site. Comparing couplings of the zero-mode to different external sectors localised at different sites is a moot issue for such class of models, since the coupling resides on a single site. The comparison between the coupling of the zero mode and the gears at the {\it same} site is a model-independent test of the localisation properties in the clockwork.

The difference between the two approaches becomes also manifest when we consider the limit to the continuum. As shown in \cite{Giudice:2016yja}, and repeated in \sect{sec:cont}, a non-trivial continuum limit requires a continuous flow of the clockwork parameter $q$ such that $q=1+ \mathcal{O}(a)$, where $a$ is the lattice spacing which eventually goes to zero. By taking general values of $q$, any recourse to the UV-completion must be from the outset abandoned.  For example, consider going to the linearly realised model through the replacement $U_j \to \Phi_j$, with the latter being a complex scalar.  Now $\Phi^q$, for $q\notin \mathbb{Z}$, can be interpreted only in an EFT context.
Thus in taking general values of $q$ only the non-linear representation of the EFT is applicable. The assumption on the external charges made in \cite{Craig:2017cda} is perfectly well-defined for integer $q$ when the complex scalar that includes the radial mode, $U_j \to \Phi_j$ is involved. However, it no longer applies when general values of $q$ are considered, which can be done only in an EFT perspective. In particular, when  $q\to 1 + \mathcal{O}(a)$, the pattern of couplings to external states becomes entirely model-dependent, as expected from the above arguments, and position-dependent and position-independent couplings are both possible for a scalar field, and both consistent with the clockworking.

Thus, in summary, much of the difference between \cite{Giudice:2016yja} and \cite{Craig:2017cda} lies on the different perspectives on the clockwork and its application purposes.  Both \cite{Giudice:2016yja} and \cite{Craig:2017cda} are focussed on the position-dependence of the zero mode coupling, however this is relative to the local cutoff in \cite{Giudice:2016yja}, and relative to other sites in \cite{Craig:2017cda}.  The work of \cite{Giudice:2016yja} was focussed entirely on constructing models based on the clockworked shift symmetry $\pi_j \to \pi_j +q^{-j}\alpha$ and the translation-independent mass terms (within an EFT perspective where it is possible to take $q\notin \mathbb{Z}$), which lead to  a hierarchy between the zero mode coupling and the cutoff.  On the other hand, the arguments of \cite{Craig:2017cda} require the basic clockwork ingredients \emph{and} a requirement on the pattern of external couplings, with a view towards generating hierarchies of zero mode couplings to different sites.  This assumption follows from locality in the charges of fields at each site.  This is based on a UV perspective, which is motivated when considering the full complex scalar that includes the radial mode, $U_j \to \Phi_j$.

\section{On the clockwork continuum limit}
\label{sec:cont}
Let us take the scalar clockwork EFT, including a coupling to an external gauge sector at the site $\ell$
\bea
{\mathcal L} & = & -\frac{1}{2}\, \sum_{j=0}^N \partial_\mu \pi_j \, \partial^\mu \pi_j- \frac{m^2}{2}  \sum_{j=0}^{N-1} c_j (\pi_j - q \pi_{j+1})^2 +{\mathcal L}_{\rm int} (\pi_j -q\pi_{j+1}) \nonumber \\
& & - \frac{1}{4g^2} G^{\mu\nu} G_{\mu\nu} + \frac{\pi_\ell}{16 \pi^2 f_0} G^{\mu\nu} \widetilde{G}_{\mu\nu}~.
\label{eq:warpvsclock1}
\eea
The parameters $c_j$ describe possible non-universality of the mass terms. Let us perform a field redefinition $\pi_j = \phi_j/q^j$.  The action becomes
\bea
{\mathcal L} & = & -\frac{1}{2}\, \sum_{j=0}^N q^{-2 j} \partial_\mu \phi_j \, \partial^\mu \phi_j- \frac{m^2}{2}  \sum_{j=0}^{N-1} q^{-2 j}c_j (\phi_{j+1}-\phi_j)^2 +{\mathcal L}'_{\rm int} (\phi_{j+1}-\phi_j) \nonumber \\
& & - \frac{1}{4g^2} G^{\mu\nu} G_{\mu\nu} + \frac{\phi_\ell}{16 \pi^2 q^{\ell} f_0} G^{\mu\nu} \widetilde{G}_{\mu\nu}~.
\label{eq:warpvsclock2}
\eea

To study the continuum limit we may now define a discretised extra coordinate $y_j =j a$ (with $j=0,...,N$), where $a=\pi R/N$ is the lattice spacing. The lattice dependence of the parameters in the potential is chosen such that~\cite{Giudice:2016yja}
\beq
m^2(a) =\frac{1}{a^2}~, ~~~~q(a)=e^{ka}~.
\eeq
Using $q^{j} =e^{ky_j}$ and taking the continuum limit $a\to 0$ (in which $(\phi_{j+1}-\phi_{j})/a \to \partial_y \phi$), we find
\bea
{\mathcal L} & = & - \frac{1}{2}\, \int dy\,  e^{-2 k y} \left[ \partial_\mu \phi \, \partial^\mu \phi + C(y) \left( \partial_y \phi \right)^2  \right] 
\nonumber \\ 
& & -\int dy \, \delta(y-y_0) \left(\frac{1}{4g^2} G^{\mu\nu} G_{\mu\nu} - \frac{e^{-k y} \phi}{16 \pi^2 f_0} G^{\mu\nu} \widetilde{G}_{\mu\nu} \right) ~.
\label{eq:lindil}
\eea

Here $y_0 = \ell a$ and $C(y)$ is the continuum limit of $c_j$. The universal case corresponds to $C(y)=1$.

The clockwork symmetry uniquely determines the 4D kinetic term, which exhibits an exponential warping. The kinetic term along the extra coordinate $y$ (and thus the gear mass spectrum) is determined by the condition of translational invariance in site space. This gives $C(y)=1$ and singles out the linear dilaton metric. Only quadratic terms have survived the continuum limit ($\lim_{a\to 0}{\mathcal L}_{\rm int} = 0$), as higher order terms scale as $ (a \partial_y \phi)^z/a^2$, where $z>2$.  

This result shows how all theories that obey the clockwork symmetry and that are described by the EFT in eqs.~(\ref{lag1})--(\ref{lag2}), independently of their interactions, flow in the continuum towards a free 5D scalar theory on a linear dilaton background. In other words, the quadratic terms of any clockwork theory entirely characterise its continuum limit.  If deconstructed, one can realise all discrete 4D clockwork models by reintroducing the operators that have vanished in the approach to the continuum limit.

The particular structure of the metric in \eq{eq:lindil} is ensured by the solution to Einstein's equations for the linear dilaton action (see refs.~\cite{Antoniadis:2011qw,Antoniadis:2001sw,Baryakhtar:2012wj,Cox:2012ee})
\beq
{\mathcal S}_S = \int d^4 x\, dy\, \sqrt{-g} \, \frac{M_5^3}{2} e^{S} \left( {\mathcal R} +   g^{MN}\partial_M S
\, \partial_N S +4k^2 \right) ~,
\label{actionstring}
\eeq
which gives $\langle S \rangle = \pm 2 k y$. Thus, the linear dilaton gives a {\it dynamical} explanation of the exponential factors appearing in the clockwork model, since the factors of $e^{-2 k y}$ are given by the couplings to the dilaton $e^S$.  

In summary, a 5D scalar in a linear dilaton background is the unique continuum limit of the prototype model in \cite{Choi:2015fiu,Kaplan:2015fuy}, as well as any model belonging to the clockwork EFT with mass terms that are translationally invariant in site space. This is why in \cite{Giudice:2016yja} this was appropriately referred to as the `continuum clockwork'. 

\subsubsection{The continuum clockwork symmetry}
As emphasised in \Sec{sec:iflifegivesyoulemmas}, in \cite{Giudice:2016yja} the symmetry structure was used as the core defining feature of the clockwork.  In the basis in which the discrete clockwork is described by \Eq{eq:warpvsclock1}, where fields at each site have canonically normalised kinetic terms, the clockwork shift symmetry is $\pi_j \to \pi_j + \alpha/q^j$, hence the wavefunction of the zero mode has an exponential profile.  In the basis of \Eq{eq:warpvsclock2}, the clockwork shift symmetry is $\phi_j \to \phi_j + \alpha$, hence the wavefunction of the zero mode is flat.

In the continuum model with a scalar on a linear dilaton background (given by \eq{eq:lindil} with $C(y)=1$), the clockwork symmetry is $\phi(y) \to \phi(y) +\alpha$, as this is the same basis as \Eq{eq:warpvsclock2}. To go to the continuum counterpart of the basis \Eq{eq:warpvsclock1}, we can define $\pi(y) = e^{-k y} \phi(y)$, so that 
the field has canonically normalised 4D kinetic terms at each position. In terms of this field, the shift symmetry is $\pi(y) \to \pi(y) + \alpha e^{-k y}$, perfectly reproducing the position-dependent clockwork symmetry, but now in continuum form. 

\section{On position dependence of couplings}
\label{sec:posdep}

We argued before that the distribution along sites of the charges of the clockwork to external sectors (and, consequently, the profile of the zero-mode couplings to external sectors)  is a model-dependent issue. We want to show here that the same is true also in the continuum limit, {\it i.e.} in the linear dilaton model. 

The linear dilaton model is determined by a shift symmetry acting on the dilaton $S \to S+ \alpha$. In the Einstein frame this is an internal symmetry. However in the Jordan frame, where we will work, this is accompanied by a corresponding global Weyl transformation $ g_{MN} \to e^{-2 \alpha/3}  g_{MN}$.  The symmetry structure of the action \Eq{actionstring} can be understood by considering $M_5$ and $k$ as dimensionful spurions that transform as $M_5 \to M_5$ and $k \to e^{\alpha/3} k$.  Thus, $k$ explicitly breaks the dilaton shift symmetry.  

The continuum clockwork scalar $\phi$ may then be coupled to the topological term of a gauge field on a brane at $y=y_0$ as
\bea
\mathcal{S} & = & \mathcal{S}_S - \int d^5 x \, e^S \sqrt{-g} \left(\frac{1}{2}  g^{MN} \partial_M \phi \partial_N \phi \right) \\
& & - \int d^5 x\,  \delta(y-y_0)  \left(\frac{\sqrt{g^{(4D)}}}{4g^2} G_{\mu\nu} G^{\mu\nu} - \frac{e^{\frac{n S}{2}} \phi}{16 \pi^2 \Lambda^{3/2}}  \, G_{\mu\nu} \widetilde{G}^{\mu\nu} \right)    ~,
\label{eq:posdepcoup}
\eea
where $\Lambda$ is a dimensionful parameter that may, like $M_5$, not break the shift symmetry ($n=0$), or breaks it like $k$ ($n=1$) or with a different spurion charge ($n\ne 0,1$).  Thus the position-dependence of the zero-mode coupling is entirely model-dependent.\footnote{In \cite{Giudice:2016yja} only a coupling at $y=0$ was considered, so there is no distinction between the different values of $n$.} For this reason, from an EFT perspective, we have reservations on using position-dependence as a defining property of the clockwork, as done in \cite{Craig:2017cda}. The model-dependence of the zero-mode coupling can be classified by two cases.

{\it a)}  If the coupling to the external gauge sector involves no explicit breaking of scale invariance ($n=0$), then the zero-mode coupling is independent of $y_0$, the position of the brane.  However, this is still a clockwork theory, as the properties of the bulk action do not depend on brane couplings, and the clockwork generation of a scale hierarchy remains.

{\it b)}  If the coupling to the external gauge sector involves explicit scale-breaking parameters ($n\neq 0$), then the zero-mode coupling  is exponentially dependent on the position. However, as we will now show, in this case the $\phi$ shift symmetry is necessarily non-compact, posing a severe limitation to interesting model building.

From an EFT perspective, either choice {\it a)} or {\it b)} is equally reasonable, unless some symmetry forbids one or the other.

\subsubsection{External charges in the continuum}
\label{sec:noncompact}
In general, the Goldstone of a compact $\text{U}(1)$ symmetry couples to external states with a strength proportional to $Q/f$, where $Q$ is the charge under the symmetry and $f$ is the decay constant.  This also applies to topological terms.  Now, if the Goldstone boson is coupled to external fields $a$ and $b$, with charge $Q_a$ and $Q_b$, then the ratio of couplings $Q_a/Q_b$ is necessarily a rational number, since the charge is quantised ($Q_a,Q_b \in \Z$).  However, if the coupling of the field $\phi$ can vary continuously in $\R$, as in the case {\it b)} above, ($Q_a/Q_b=e^{nk (y_b-y_a)}$), then the ratio of charges to external fields located at generic different positions $y_a$ and $y_b$
can take any real value ($Q_a/Q_b \in \R$). Since there is no charge quantisation, the shift symmetry is necessarily non-compact.

To put this in more direct terms, a spontaneously broken compact symmetry has a discrete gauge symmetry, such that for $\phi \to \phi + 2 \pi n f$, the theory is physically identical. If the charges can take also irrational values, then there is no discrete gauge symmetry for the scalar and this cannot be a Goldstone.  Note that this breaking comes entirely from the brane couplings, thus the bulk action, including the kinetic terms, may still possess this symmetry.  Of course, if the external sector is coupled only to one site, the symmetry can be compact. However, in this case there is no notion of the profile of the external charges, and no notion of position-dependence, thus the point of position-dependence becomes moot.

The non-compact possibility is less appealing for model-building applications. Moreover, it has been argued that no non-compact symmetries can arise in quantum gravity, whether global or gauged~\cite{Banks:2010zn}. Even within field theory, if the goal is to justify large ratios of couplings to external states with non-compact symmetries, one might as well just sidestep the entire clockwork machinery since infinite field excursions are already consistent with a non-compact symmetry.

\section{On the continuum proposal of \cite{Craig:2017cda}}
\label{sec:theirSec}

We have shown that the continuum limit of the clockwork EFT is unique and is described by the linear dilaton model. However, the authors of \cite{Craig:2017cda} suggest a different model as the appropriate continuum limit. Here we show that the two models are identical, up to a field redefinition.

Let us take the continuum clockwork model of a massless scalar in the linear dilaton background, \Eq{eq:lindil} for $C(y)=1$ and the choice of coupling to the external sector corresponding to $n=1$, in the notations of \eq{eq:posdepcoup}. After the field redefinition $\phi = e^{k y} \pi$, the Lagrangian becomes
\bea
{\mathcal L} & = & -\frac{1}{2}\, \int dy\, \left[ \partial_\mu \pi \, \partial^\mu \pi + \left( \partial_y \pi \right)^2 + k^2 \pi^2 + k \partial_y \pi^2 \right] \\
& & -\int dy\, \delta(y-y_0) \left( \frac{1}{4g^2} G^{\mu\nu} G_{\mu\nu} - \frac{\pi}{16 \pi^2 f_0} G^{\mu\nu} \widetilde{G}_{\mu\nu} \right) ~,
\eea
which, using the identity
\be
\int_0^{\pi R} dy\,  \partial_y F(y) = 2 \int_0^{\pi R} dy\, F(y) \left[ \delta(y-\pi R) - \delta(y)  \right] ~,
\ee
can be written as
\bea
{\mathcal L} & = & -\frac{1}{2}\, \int dy\, \bigg[ \partial_\mu \pi \, \partial^\mu \pi + \left( \partial_y \pi \right)^2 + \pi^2 \left( k^2 +2k\, \delta(y-\pi R) -2k\, \delta(y)\right)  \bigg] \\
& & -\int dy\, \delta(y-y_0) \left( \frac{1}{4g^2} G^{\mu\nu} G_{\mu\nu} - \frac{\pi}{16 \pi^2 f_0} G^{\mu\nu} \widetilde{G}_{\mu\nu} \right) ~.
\eea

The continuum limit of the clockwork proposed in \cite{Craig:2017cda} is described by the action
\bea
{\mathcal S} & = & - \frac{1}{2}\,\int d^4 x\,  dy \,  \bigg[  \partial_\mu \phi \, \partial^\mu \phi +  (\partial_y \phi)^2  + \phi^2 \left(M_\phi^2+  \tilde{m}_\phi \, \delta(y) - \tilde{m}_\phi \, \delta(y-\pi R)   \right) \bigg] \nonumber \\
& & -\int d^4 x\,  dy \, \delta(y-y_0) \left(\frac{1}{4g^2} G^{\mu\nu} G_{\mu\nu} - \frac{\phi}{16 \pi^2 \Lambda^{3/2}}\,  G^{\mu\nu} \widetilde{G}_{\mu\nu} \right) \bigg]  ~,
\label{uffaactionthem}
\eea
where, to obtain a massless scalar mode, one must require $\tilde{m}_\phi = \pm 2 \sqrt{M_\phi^2}$.  Thus, the continuum limit proposed in \cite{Craig:2017cda} is simply the scalar theory on a linear dilaton background.  

As explained in \Sec{sec:posdep}, the choice made in \cite{Craig:2017cda} for the coupling to the external sector ($n=1$) corresponds to a non-compact shift symmetry, which is not a favourable situation for model building, nor for considering super-Planckian field excursions or addressing the WGC. 

The authors of \cite{Craig:2017cda} also provide a comparison of discretised versions of the linear dilaton model and their continuum model.  As the two theories are the same, the authors are simply demonstrating the existence of different lattice artifacts ($1/N$ corrections) emerging from two different latticisations of the same theory.

\section{On the localisation of the zero mode in the continuum clockwork}
\label{sec:local}
Unlike the claim in \cite{Craig:2017cda}, it was found in \cite{Giudice:2016yja} that the zero mode of the linear dilaton is physically localised. Let us review the argument made in \cite{Giudice:2016yja}.

Consider the action of a 5D scalar $\phi$ on a linear dilaton background and expand the field as
\beq
\phi (x,y) = \sum_{n=0}^\infty \frac{{\tilde \phi}_n (x)\, \psi_n (y)}{\sqrt{\pi R}} ~.
\eeq
The requirement that the original action is equal to the sum of the actions for the 4D fields ${\tilde \phi}_n$, with canonical kinetic terms, imposes the orthonormal conditions 
\beq
\int_{-R\pi}^{R\pi}\frac{dy}{\pi R}\, e^{2k|y|}\psi_n(y) \psi_m(y) = \delta_{nm}~.
\eeq
Thus, the norm of the zero mode and the gears are given by
\bea
|| \psi_0 ||^2 &=& \int_{-\pi R}^{\pi R} dy\, \frac{k\, e^{2k|y|}}{e^{2k\pi R}-1} 
\label{eigen1}\\
|| \psi_n ||^2 &=& \int_{-\pi R}^{\pi R} \frac{dy}{2\pi R} \left[ 1+ \frac{n^2 -k^2R^2}{n^2 +k^2R^2} \cos \frac{2ny}{R} + \frac{2nR}{n^2 +k^2R^2} \sin \frac{2n|y|}{R} \right]  ~~~ n \in \mathbb{N} ~.
\label{eigen2}
\eea
The integrands of eqs.~(\ref{eigen1}) and (\ref{eigen2}) have a quantum interpretation of probability densities  $d || \psi_n ||^2/dy$. Their expressions show the exponential localisation, in proper distance, of the zero mode at one end of the orbifold, and the roughly uniform distribution (up to a periodic modulation) of the gears.  These distributions were shown in \cite{Giudice:2016yja} to be the continuum limit of the discrete case.  

More generally, note that the 4D mass of a KK mode from a 5D massless field is its momentum in the $y$-direction: $E^2-{ \vec p}^{\, 2} = p_5^2$. Thus, the zero mode wavefunction ($\phi_0$) is flat, as it carries no $y$-momentum: $\partial_y \phi_0$.  However, this does not imply that the zero mode is not physically localised, as is well known for the RS graviton~\cite{Randall:1999ee}, which also has a flat wavefunction, but is localised at the UV brane.

The localisation of the zero mode in coordinate space and the delocalisation of the gears translate into a physical observable, when we consider the ratio of couplings to an external sector at a given location. Such ratio, in the case of the linear dilaton model, exhibits the characteristic clockwork property of being of order unity at one orbifold fixed point and exponentially suppressed on the other brane.

\section{On warping and clockworking}
\label{sec:warpvsclock}

Let us compare eqs.~(\ref{eq:warpvsclock1}) and (\ref{eq:warpvsclock2}), which describe the scalar clockwork EFT in two field bases. 
Equation~(\ref{eq:warpvsclock1}) exhibits the familiar structure of clockworking, with canonical kinetic terms and nearest-neighbour mass terms skewed by the parameter $q$. Equation~(\ref{eq:warpvsclock2}) exhibits the properties of warping, defined as exponentially varying wave-function renormalisations. As previously discussed, both bases have their continuum counterparts.

The two Lagrangians describe the same QFT, as they are related by a simple field redefinition, and thus reveal the intimate relation between clockworking and warping. However, while the clockwork symmetry uniquely determines the warping factor ($q^{-2j}$ in front of the kinetic term), the mass spectrum needs a further requirement, here chosen as translational invariance in site space. This highlights why warping from clockworking is not the same as in RS. As pointed out in \cite{Giudice:2016yja}, the difference lies in the mass terms, which are exponentially varying in RS (in the canonical basis) and universal in the clockwork. This difference leads to distinct mass spectra and the characteristic mass gap of the clockwork gears.

The authors of \cite{Craig:2017cda} make a conceptual distinction between the two theories in eqs.~(\ref{eq:warpvsclock1}) and (\ref{eq:warpvsclock2}). While \eq{eq:warpvsclock1} is classified as clockwork, \eq{eq:warpvsclock2} is not, because of apparently ad-hoc exponentially varying parameters. The claim is that the distinction becomes important when the EFT is matched to a fundamental theory in the UV. While it is certainly true that a particular field basis can be more revealing about the symmetries or properties of the underlying fundamental theory, and thus more appropriate for a straightforward UV matching, it is difficult for us to take the point of view that two theories related by a field redefinition (thus being the same QFT) should be classified differently. For this reason, we think it is appropriate to call a theory clockwork, whether described by the Lagrangian in \eq{eq:warpvsclock1} or (\ref{eq:warpvsclock2}).

Note that in \Eq{eq:warpvsclock1} the shift symmetry is clockworked $\pi_j \to \pi_j +q^{-j}\alpha$, leading to an exponential wavefunction for the zero mode.  Yet in \Eq{eq:warpvsclock2} the shift symmetry is symmetrically shared $\phi_j \to \phi_j +\alpha$, meaning that in this basis the wavefunction of the zero mode is flat.  The resulting physics is the same in both cases, as the theories are the same.

As shown in \Sec{sec:cont}, the continuum limit of the scalar clockwork is given by
\beq
{\mathcal L} = -\frac{1}{2}\, \int dy\, e^{-2 k y} \, \partial_M \phi \partial^M \phi ~.
\label{eq:LDM}
\eeq
Here the warp factor $e^{-2ky}$ is the continuum analog of $q^{-2j}$ in the discrete model.  Once again the action looks very different than the original clockwork, however it remains that $e^{-2 k y}$ is no more an exponential parameter entered by hand than is $q$ in \Eq{eq:warpvsclock1}.  As a result, the physical mechanism localising the zero mode and generating the mass spectrum in \Eq{eq:LDM} is precisely the same mechanism as in \Eq{eq:warpvsclock1}.  

%Let us now go beyond the quadratic level and consider interactions.  One interaction we may add is
%\beq
%{\mathcal L} = \int dy\, e^{-2 k y} \left[ -\frac{1}{2}\, \partial_M \phi \partial^M \phi
%+\frac{1}{\Lambda^5} \left( \partial_M \phi \partial^M \phi \right)^2 \right] ~ .
%\label{eq:LDMint}
%\eeq
%Deconstructing this action leads to the theory
%\bea
%&{\mathcal L}&  =  -\frac{1}{2}\, \sum_{j=0}^N \,
%\partial_\mu \pi_j \, \partial^\mu \pi_j  -\frac{1}{2}\,  \sum_{j=0}^{N-1} m^2 (\pi_{j} -q \pi_{j+1})^2  \\
%&  +&\frac{1}{\widetilde{\Lambda}^4} \left[ \sum_{j=0}^N \,
%q^{2 j} \left( \partial^\mu \pi_j \right)^4 + 2 \sum_{j=0}^{N-1} q^{2 j} m^2 (\partial^\mu \pi_j)^2 (\pi_{j} -q \pi_{j+1})^2+ \sum_{j=0}^{N-1} q^{2 j} m^4 (\pi_{j} -q \pi_{j+1})^4 \right] ~ .\nonumber
%\label{eq:warpvsclockint}
%\eea
%The new interaction terms are site-dependent, and vary exponentially from one side of the chain to another.  However the model has not ceased to be a clockwork model, as the underlying clockwork mechanism is the same as in \Eq{eq:warpvsclock1}, as is the quadratic action that specifies the localisation and mass spectrum of all modes.

\section{On clockwork gravity}
\label{sec:gravity}
The linear dilaton model has been discussed comprehensively in \cite{Antoniadis:2011qw,Antoniadis:2001sw,Baryakhtar:2012wj,Cox:2012ee}, and with regard to its continuum clockworking properties in \cite{Giudice:2016yja}.  When discretised, the gravitational part of the action is
\be
\mathcal{S} = M_P^2 \sum_{j} \int d^4 x \, e^{S_j} \sqrt{-g_j} \left[ R_j + \frac{m^2}{4}\left( \left(g_{j,\mu\nu} (g^{\mu\nu}_{j+1}-g^{\mu\nu}_{j})\right)^2 + (g^{\mu\nu}_{j+1}-g^{\mu\nu}_{j})^2 \right) \right] ~,
\label{eq:discrete}
\ee
where $m^2$ and $M_P$ are the same everywhere, as the initial action was translation invariant.  In the continuum the equation of motion is $\partial_y S =-2k$, thus in the discrete this solution corresponds to $S_{j+1}-S_j =-2ka$. Using the dictionary between continuum and discrete $ka=\ln q$, this corresponds to $S_j = \ln q^{-2j}$.  The action becomes
\be
\mathcal{S} = M_P^2 \sum_{j} \int d^4 x \, q^{-2 j} \sqrt{g_j}  \left[ R_j + \frac{m^2}{4}\left( \left(g_{j,\mu\nu} (g^{\mu\nu}_{j+1}-g^{\mu\nu}_{j})\right)^2 + (g^{\mu\nu}_{j+1}-g^{\mu\nu}_{j})^2 \right)\right]  ~,
\label{pippo}
\ee
where the individual Planck scales are warped $q^{-j} M_P = M_j$, and the diffeomorphism invariance is shared symmetrically amongst the sites.  This is analogous to the warped basis for the original clockwork model, \Eq{eq:warpvsclock2} where there is a $q^{-2 j}$ warp factor, no more an exponential parameter entered by hand than as in \Eq{eq:warpvsclock2}, and the symmetry is shared symmetrically, as expected.

Expanding the metric to graviton fluctuations $g^j_{\mu\nu}= \eta^{\mu\nu} + \frac{2}{M_j} h^j_{\mu\nu}$ the diffeomorphism invariance at each site is realised linearly as a translation invariance at each site, by the gauge symmetry
\be
h^j_{\mu\nu} \to h^j_{\mu\nu} + \partial_\mu A^j_\nu + \partial_\nu A^j_\mu +\frac{2}{M_j} f^{\rho\sigma}_{\mu\nu} \partial_\rho A^{j,\alpha} h^j_{\alpha \sigma} ~ ,
\ee
where $f^{\rho\sigma}_{\mu\nu} = \delta^\rho_\mu \delta^\sigma_\nu + \delta^\rho_\nu \delta^\sigma_\mu$.\footnote{In \cite{Giudice:2016yja} only the lowest order terms in the graviton perturbation were shown, for simplicity.}  However, the nearest-neighbour interactions break these different diffeomorphism invariances to the clockworked subgroup $A^j_\nu = q^{-j} \tilde{A}_\nu$, such that the remaining symmetry is now shared asymmetrically amongst sites
\be
h^j_{\mu\nu} \to h^j_{\mu\nu} +q^{-j} \left( \partial_\mu \tilde{A}_\nu + \partial_\nu \tilde{A}_\mu +\frac{2}{M_j} f^{\rho\sigma}_{\mu\nu} \partial_\rho \tilde{A}^{\alpha} h^j_{\alpha \sigma} \right)~ .
\ee
This clockworked symmetry enforces that the zero mode is localised exponentially relative to the graviton at site $j$
\be
h_j^{\mu\nu} = \frac{1}{q^j} \tilde{h}_0^{\mu\nu} +...
\label{eq:overlap}
\ee
where the ellipsis contains the heavy graviton gears.   Note that the Planck scale varies across the sites, such that $M_j = q^{-j} M_P$, hence one may equivalently write this transformation as
\be
h^j_{\mu\nu} \to h^j_{\mu\nu} +q^{-j} \left( \partial_\mu \tilde{A}_\nu + \partial_\nu \tilde{A}_\mu \right) +\frac{2}{M_P} f^{\rho\sigma}_{\mu\nu} \partial_\rho \tilde{A}^{\alpha} h^j_{\alpha \sigma} ~ .
\ee
The higher order terms automatically preserve the 4D diffeomorphism invariance, as the action was invariant from the beginning.  Expanding the metric to quadratic order in the linear perturbations (which we may take in transverse-traceless gauge), gives the discrete multi-graviton masses terms of \cite{Giudice:2016yja}, 
\be
\mathcal{S}_{int} = -\sum_{j} \int d^4 x \frac{m^2}{2} \left[ \left( h^j_{\mu\nu} - q h^{j+1}_{\mu\nu} \right)^2 + \mathcal{O}(h^3) \right]  ~.
\ee
The interaction terms follow directly from the expansion of \Eq{pippo}.  In the continuum case the same behaviour remains, with the replacement $q^{-j} \to e^{-k y}$.  

Multi-graviton models suffer from numerous technical subtleties, see e.g. \cite{deRham:2014zqa} for a thorough review. However this multi-graviton scenario has no worse behaviour than other multi-gravity model, while, as in the continuum, the multi-site dilatons play a crucial role in the consistency of the theory.

\subsubsection{External fields}
To exhibit the effect of coupling the graviton clockwork to a localised external sector, let us consider the simple case of a scalar $H$ residing at a single site $\ell$.
The coupling of the graviton to the scalar is
\be
\mathcal{L}_\ell = \frac{1}{M_\ell} h_\ell^{\mu\nu} \left[ \partial_\mu H \partial_\nu H -\frac{1}{2} \eta_{\mu\nu} (\partial_\rho H \partial^\rho H + \kappa^2 M_\ell^2 H^2)  \right] ~,
\ee
where $\kappa$ is a parameter that measures the scalar mass in natural units. The local cutoff of the interaction is $M_\ell$, hence we should expect the scalar mass to be ${\mathcal O} (M_\ell)$ and $\kappa$ of order unity.  The overlap between the graviton zero mode and the graviton at site $\ell$ is given in \Eq{eq:overlap}, thus the interaction of the zero mode becomes
\be
\mathcal{L}_\ell = \frac{1}{q^\ell M_\ell} \tilde{h}_0^{\mu\nu} \left[ \partial_\mu H \partial_\nu H -\frac{1}{2} \eta_{\mu\nu} (\partial_\rho H \partial^\rho H + \kappa^2 M_\ell^2 H^2)  \right]  ~.
\ee
The zero mode is indeed a factor $1/q^\ell$ more weakly coupled to matter than the gears or, equivalently, than what suggested by the local cutoff.

On any site, the only physical observables are the ratios of scales.  For example, if the scalar has an $\mathcal{O}(1)$ self coupling and obtains a vev $v_H \sim M_j$, then the hierarchy between the vev and the graviton interaction scale is $v_H / M_P \sim q^{-\ell}$.   At $\ell=0$ there is no hierarchy, whereas at $\ell=N$ the hierarchy is exponential.  

Since the Planck scale at each site is $M_\ell =M_P/q^{\ell}$, indeed the strength of the graviton interaction is the same at all $\ell$, which is the objection of \cite{Craig:2017cda}.  However, this is simply what diffeomorphism invariance theory has handed us as the only allowed possibility for couplings to external states.  The graviton is still clockworked, as demonstrated in \Eq{eq:overlap}.

\subsubsection{Exponential input parameters}
The clockwork is usually presented in terms of $N\! +\! 1$ different $U(1)_j$ symmetries, under which each complex scalar $\Phi_j$ carries unit charge, explicitly broken by nearest neighbour spurions to a single $U(1)_{CW}$, where the spurions carry charge $(1,-q)$ under $U(1)_j \times U(1)_{j+1}$.  This is a useful bookkeeping tool, however with multiple $U(1)$ symmetries one can always rotate the $U(1)$ generators to choose a different basis that explicitly separates out continuous from approximate symmetries $\prod^{N}_{j=0} U(1)_j = U(1)_{CW} \times \prod^N_{j=1} \widetilde{U}(1)_j$.  The latter comprise only the explicitly broken symmetries, and the spurions carry charge only under $\widetilde{U}(1)_j$.

In this basis, the complex scalar fields $\Phi_j$ have an explicit hierarchy of charges $Q^{CW}_j = 1,q^{-1},...,q^{-N}$ under the continuous $U(1)_{CW}$ symmetry.  Furthermore, any operator they couple to must also have this charge $Q^{CW}_j = 1/q^j$, which follows directly from the input charges $Q^{CW}_j = 1/q^j$.  In other words, as a result of the input charges, the zero-mode decay constant at each site is $f_j \propto f_0 q^j$.

The background value of the dilaton field $e^S$ is related to the string coupling as $e^S \propto 1/g_s^2$, where the proportionality involves the volume of the extra dimensions.  Dynamically, as a solution of Einstein's equations, the dilaton, and hence the string coupling, takes the value at each site $g_{s_j} = g_{s_0} q^j$.  This is in direct analogy with the clockwork scalar model, and is no more of an exponential hierarchy of input parameters as the choice of charges $Q^{CW}_j = 1,q^{-1},...,q^{-N}$ was in the clockwork model.  In some sense, it is better motivated as it arises dynamically.  Since this $q^j$ factor is the same as in going from \eq{eq:warpvsclock1} to \eq{eq:warpvsclock2}, we see no reason to deem either theory to exhibit more of an exponential hierarchy of input parameters than the other.  

\section{Summary}
\label{sec:summary}

We summarise here the various points that we have made in this note regarding the criticism of \cite{Giudice:2016yja} put forth by the authors of \cite{Craig:2017cda}.

\begin{itemize}

\item The goal of \cite{Giudice:2016yja} was to show that the mechanism proposed in \cite{Choi:2015fiu,Kaplan:2015fuy} can be generalised to a variety of models with interesting phenomenological applications. These generalisations were designed to retain the original symmetry structure (clockworked symmetry and translational invariance in site space), which was taken as the defining feature of the clockwork. This symmetry structure leads to properties that can be viewed as hallmarks of the clockwork: the localisation of the zero mode, the characteristic mass spectrum of the gears, and the exponential hierarchy between the zero-mode interaction scale and the cutoff scale (measured by the gear interactions). All models presented in \cite{Giudice:2016yja} exhibit this symmetry structure and the corresponding properties.

\item In the models proposed in \cite{Giudice:2016yja} the coupling of the clockwork to the external sector resides at a {\it single} site, because this is the crucial ingredient necessary to obtain the desired phenomenological features. These models were {\it not} intended to address issues like transplanckian field excursions or the WGC. 

\item Differences between \cite{Giudice:2016yja} and \cite{Craig:2017cda} arise from different perspectives on the clockwork and its applications.  Both \cite{Giudice:2016yja} and \cite{Craig:2017cda} are focussed on the position-dependence of the zero mode coupling, however this is relative to the local cutoff in \cite{Giudice:2016yja}, and relative to other sites in \cite{Craig:2017cda}. The definition of clockwork employed in \cite{Craig:2017cda} is appropriate to address various UV-related features of the zero mode, but is not useful in the domain of the models considered in  \cite{Giudice:2016yja}. In the EFT approach of \cite{Giudice:2016yja}, the comparison of  the clockwork couplings to external sectors at {\it different} sites is a model-dependent issue that cannot be settled without UV knowledge. Moreover, the issue of comparing couplings at different sites is moot for models in which the coupling is localised at a single site.  On the other hand, the comparison between the interaction scales of the zero-mode and the gears (cutoff) at the same site corresponds to a physical observable, well-defined even in a general EFT context with no reference to the UV. The site-dependence of this ratio of scales is
a genuine test of the clockwork localisation properties.

\item We have reaffirmed the linear dilaton model as the continuum limit of the discrete scalar clockwork. More generally, we have shown that the unique continuum limit of the model in \cite{Choi:2015fiu,Kaplan:2015fuy}, as well as of any model sharing the same symmetry structure and belonging to the class of clockwork EFT, is given by a 5D free massless scalar field on a linear dilaton background.  As a result, any massless field placed in this background leads to the characteristic clockwork properties coming from its symmetry structure: zero-mode localisation, gear mass spectrum, and position-dependent hierarchy between the zero mode coupling and the cutoff of the interaction. In particular, the zero mode in the linear dilaton model satisfies the localisation properties of the clockwork, exactly in the same way as in discrete models. 

\item The model-dependence of the couplings to external sectors along the sites -- previously encountered in the discrete -- is also found in the continuum. In the context of the linear dilaton, this model-dependence is related to the (a-priori unknown) properties of the interactions under the dilaton shift symmetry. Insisting on position-dependent zero-mode couplings leads necessarily to a non-compact symmetry in the continuum, restricting model-building opportunities. 

\item The model proposed in \cite{Craig:2017cda} as the successful continuum clockwork is identical to the linear dilaton model, after a field redefinition.

\item Based on UV considerations, the authors of \cite{Craig:2017cda} argue to distinguish their continuum model from the linear dilaton. While different choices of basis can be more or less convenient for UV matching, the physical content of a quantum field theory does not change depending on the field basis. For this reason, we prefer not to classify with different names EFTs that are identical, up to a field redefinition, and we collectively call them `clockwork'.

\item The intimate relation between clockworking and warping, both in the discrete and the continuum, is shown by a simple field redefinition. Depending on the choice of basis, one can view the same theory as a clockwork with canonical kinetic terms and nearest-neighbour skewed mass terms or as a warped model with exponentially varying wave-function renormalisations and symmetric nearest-neighbour mass terms. Moreover, the mass universality condition singles out the characteristic clockwork mass spectrum, thus distinguishing clockworking from warping in RS models.

\item Once the effect of the dilaton is properly taken into account, the discretised version of linear dilaton theory behaves like a discrete clockwork gravity model, sharing all the characteristic features of the clockwork mechanism. The dilaton background, and its interpretation as the string coupling, gives a dynamical justification of the exponentially varying Planck masses from site to site.

\end{itemize}

\bigskip
\subsubsection*{Acknowledgements}
We thank Brando Bellazzini, Yevgeny Kats, Riccardo Rattazzi, Matt Strassler, Riccardo Torre, and Alfredo Urbano for useful discussions and Nathaniel Craig, Isabel Garcia-Garcia, and David Sutherland for constructive conversations and clarifications.

\end{document}